\documentstyle[12pt]{article}
\newcommand{\me}{\mbox{$\mu_{ex}$}}
\newcounter{abc}

\begin{document}

\title{Accurate Measurements of the Chemical Potential \\of Polymeric
Systems by Monte-Carlo Simulation}

\author{N. B. Wilding and M. M\"{u}ller \\
{\small Institut f\"{u}r Physik, Johannes Gutenberg Universit\"{a}t,}\\
{\small Staudinger Weg 7, D-55099 Mainz, Germany}}
\date{}
\setcounter{page}{0}
\maketitle

\begin{abstract}

We present a new Monte-Carlo method for estimating the chemical
potential of model polymer systems. The method is based upon the
gradual insertion of a penetrable `ghost' polymer into the system and
is effective for large chain lengths and at high densities. Insertion
of the ghost chain is facilitated by use of an expanded ensemble in
which weighted transitions are permitted between states characterising
the strength of the excluded volume and thermal interactions
experienced by the ghost chain. We discuss the implementation and
optimisation of the method within the framework of the bond
fluctuation model, and demonstrate its precision by a calculation of
the finite-size corrections to the chemical potential.

\end{abstract}

\thispagestyle{empty}
\begin{center}
PACS numbers 82.20.Wt, 05.70.-a,64.70.Fx, 82.60.Lf
\end{center}
\newpage

\section{Introduction}

Measurements of chemical potential by computer simulation are an
important tool for determining the phase behaviour of model
systems. For many years the standard technique for achieving this in
atomic and simple molecular systems, has been the Widom particle
insertion method \cite{WIDOM}. This method involves repeatedly
sampling the potential $U_t$ experienced by a single particle when it
is added at random to the system of interest. The excess chemical
potential \me\ is then given by

\begin{equation}
\beta\me\ =-\ln<\exp (-\beta U_t)>
\label{eq:muex}
\end{equation}

For polymeric systems, however, the basic Widom method cannot be
employed for obtaining estimates of \me . The extended structure of
polymers renders the method unsuitable for all but the shortest chains
and lowest densities, there being a prohibitively small probability of
inserting a polymer at random without violating excluded volume
restrictions. Clearly therefore different approaches are called for,
and indeed several new techniques for measuring the chemical potential
of model polymer systems have recently been developed \cite{KUMAR}. We
describe them in turn.

A method that ameliorates some of the difficulties associated with
random chain insertion is the so-called configurational bias Monte
Carlo method, first proposed by Siepmann \cite{SIEPMANN1} and later
generalised by Frenkel and coworkers \cite{FRENKEL}, and de Pablo and
coworkers \cite{DEPABLO}. This technique utilises a biased insertion
technique to `grow' a polymer of favourable energy into a frozen
snapshot of the system, thereby extending the range over which the
Widom insertion technique is viable. A book-keeping scheme maintains a
record of the statistical bias associated with choosing energetically
favourable chain conformations, and this bias is subsequently removed
when the chemical potential is calculated. The method is also amenable
for use with Gibbs ensemble simulations, facilitating direct study of
polymer phase equilibria \cite{MOOJI,SIEPMANN2}.  However in practice,
the utility of the configurational bias approach is limited since it
tends to break down for longer chain lengths ($\ge 20$ monomers) and
higher densities ($\rho \ge 0.6$).

Another technique for obtaining \me\ is the Chain Increment method of
Kumar and coworkers \cite{KUMAR,KUMAR1,KUMAR2}. This method hinges on the
observation that the chemical potential of a chain of length $N$ is
equivalent to the sequential sum of the incremental chemical
potentials of all successively shorter chains $N-1, N-2,...$. Since
the incremental chemical potential seems to become largely independent
of chain length for longer chains ($N \ge 15$) \cite{SZLEIFER}, it is
possible to estimate the total chemical potential for a chain of
arbitrary length by calculating the incremental chemical potential
associated with adding one monomer to a chain end, and multiplying by
the chain length.  Separate sets of simulations of short chains
($N<15$) may then be performed to provide corrections associated with
the small-$N$ dependence of the incremental chemical potential. This
method thus allows estimates of the chemical potential for arbitrary
chain lengths, and indeed has facilitated the calculation of the
liquid-vapour phase diagram for a bead-spring polymer model
\cite{SHENG}.

Recently M\"{u}ller and Paul have developed a novel approach to
calculating the chemical potential of polymer chains
\cite{MUELLER}. Their method involves a thermodynamic integration over
the excluded volume interaction of a penetrable `ghost' chain immersed
in a system of `real' polymers. A control parameter $\lambda$ tunes
the strength of this interaction such that for $\lambda=1$ the ghost
decouples from the normal polymers, while for $\lambda=0$ it behaves
as a normal polymer. Separate simulations are performed at each of
several $\lambda$ values in the range (0,1), and thermodynamic
integration with respect to $\lambda$, aided by multi-histogram
extrapolation techniques, are used to obtain the chemical potential.
The method was successful in obtaining \me\ for lattice polymers up
to $N=80$ monomers and at melt densities ($\rho=0.5$).

In this paper we present a further method for calculating the chemical
potential of polymer chains that is effective both at large chain
lengths and high densities. The method is efficient, requiring only a
single simulation, and exact in the sense that no approximations are
involved. In the following sections we describe the method and discuss
its implementation and optimisation.

\section{Method}

\subsection{Background}

The strategy of the present method is similar in spirit to that of
M\"{u}ller and Paul \cite{MUELLER}. We consider a system of volume
$V=L^d$ containing $n$ polymers of length $N$, plus a single penetrable
ghost polymer. The monomers of the normal polymers experience an
attractive short-range interaction $J({\underline r})$ and are subject
to strict excluded volume restrictions. An intra-molecular bond
potential $U({\underline r})$ controls the bond lengths and angles
between neighbouring monomers of the chain.

The ghost polymer is subject to the same intra-molecular bond
potential as the normal polymers and its monomers must not overlap
with one another. However, the excluded volume restriction between
ghost monomers and normal monomers is now replaced by an {\em
athermal} effective repulsive potential $K=-\frac{1}{\beta }ln
\lambda$, where $\lambda$ is a parameter, variable in the range
$(0,1)$, that controls the strength of the repulsion. The magnitude of
the attractive thermal interaction between ghost monomers and normal
monomers is also modified by a factor $(1-\lambda)$.

For a given choice of $\lambda$, the canonical partition function of
the system is given by

\begin{equation}
Z(n,N,V,\lambda ) = \prod_{i=1}^{(n+1)N} \left \{ \int dr_i \right \}
e^{-\beta [E_t(\{{\underline r}\})+E_a(\{{\underline r}\})]}
\label{eq:Z}
\end{equation}
where the integration is over all configurations of the monomers
$i$, subject to excluded volume restrictions.

The thermal part of the configurational energy $E_t(\{{\underline
r}\})$ is defined by

\begin{equation}
E_t(\{{\underline r}\})=\sum_{(j,k)} J(|{\underline
r_j}-{\underline r_k}|) \delta_{\alpha_j\alpha_k} + (1-\lambda
)\sum_{(j,k)} J(|{\underline r_j}-{\underline r_k}|)
(1-\delta_{\alpha_j\alpha_k}) +\sum_{(l,m)}U({\underline r_l}-{\underline r_m})
\end{equation}
where $j$ and $k$ run over all monomers, while $l$ and $m$ run over all
neighbouring pairs of monomers on the same chain. We assign $\alpha_j=1$ for a
normal monomer and $\alpha_j=0$ for a ghost monomer.

The athermal part of the configurational energy $E_a(\{{\underline
r}\})$ is given by

\begin{equation}
E_a(\{{\underline r}\})=-\frac{1}{\beta }N_o(\{{\underline r}\})ln \lambda
\end{equation}
where $N_o(\{{\underline r}\})$ is the total number of overlaps between the
`excluded volume' shell of ghost monomers and those of normal
monomers.

Clearly, by tuning $\lambda$ one is able to interpolate smoothly between
two special cases:

\begin{itemize}

\item $\lambda=0$, for which the ghost behaves as a normal polymer

\item $\lambda=1$, for which the ghost decouples from the polymer system.
\end{itemize}
The ratio of the partition function at these limits defines the
excess chemical potential:

\begin{equation}
\exp(\me ) =\frac{Z(n,N,V,1)}{Z(n,N,V,0)}
\label{eq:Zrat}
\end{equation}

Following the recently proposed method of expanded ensembles
\cite{LYUBARTSEV,SIMTEMP}, we now consider an ensemble
 that facilitates direct measurement of the ratio of the partition
functions appearing in equation~\ref{eq:Zrat}. The ensemble is defined
by the partition function:

\begin{equation}
\Omega=\sum_{m=0}^{M-1} Z_m(n,N,V,\lambda_m)\exp(\eta_m)
\label{eq:Ze}
\end{equation}
where $0 \leq m \leq (M-1)$ indexes a set of $M$ $\lambda$-states spanning
the range $(0,1)$, with $\lambda_0=0$ and $\lambda_{M-1}=1$. The
$Z_m$ are the canonical partition functions defined in
equation~\ref{eq:Z} with $\lambda=\lambda_m$, while the $\eta_m$
are positive pre-weighting factors the significance of which will be
described.

To realise a Monte-Carlo simulation within this expanded ensemble, it
is necessary to allow for two types of MC steps. Firstly there are the
usual trial monomer displacements at constant $\lambda$. Secondly
there are trial transitions between neighbouring $\lambda$-states, the
spatial coordinates of all polymers remaining fixed. These
$\lambda$-transitions are accepted or rejected according to some
criterion satisfying detailed balance, such as metropolis:

\begin{equation}
P_a(n\rightarrow m)=min (1,\exp [ N_o(\ln\lambda_m - \ln
\lambda_n )+\beta(\lambda_n-\lambda_m)E_G+ \eta_m-\eta_n])
\label{eq:pacc}
\end{equation}
where $E_G=\sum_{(j,k)} J(|{\underline r_j}-{\underline r_k}|)
(1-\delta_{\alpha_j\alpha_k})$ is the interaction energy of the ghost
polymer with the other polymers. It should be stressed, however, that
these $\lambda$-transitions do not drive the system out of equilibrium
as they would if we were considering a simple canonical
ensemble. Within the expanded ensemble framework, both types of
moves---monomer displacements {\em and} $\lambda$-transitions---are
necessary to bring the system into equilibrium.

In the course of the expanded ensemble simulation a histogram
$p(\lambda_m)$ can be accumulated describing the probability with
which each of the $M$ $\lambda$-states is visited. Formally this
probability distribution takes the form:

\begin{equation}
p(\lambda_m)=\frac{Z_m}{\Omega }\exp (\eta_m)
\label{eq:pm}
\end{equation}
{}From which it follows that

\begin{equation}
\frac{p(\lambda_{M-1})}{p(\lambda_0)}=
\frac{Z_{M-1}\exp (\eta_{M-1})}{Z_0\exp (\eta_0)}
\label{eq:prat}
\end{equation}
The excess chemical potential is then calculable as:

\begin{equation}
\me = \ln\left [ \frac{\tilde p(\lambda_{M-1})}{\tilde p(\lambda_0)}\right ]
\label{eq:ptilrat}
\end{equation}
where
\begin{equation}
\tilde p(\lambda_m)\equiv p(\lambda_m)\exp(-\eta_m)
\label{eq:ptilde}
\end{equation}

Now, the principal difficulty in implementing such a scheme, is that
the ratio of the partition functions $Z_{M-1}/Z_0$ appearing in
equation~\ref{eq:prat} can (even for short chains) span very many
orders of magnitude. In order to obtain a histogram $p(\lambda_m)$
having counts for all $m$, it is therefore necessary to employ a
pre-weighting scheme that will encourage the simulation to sample
$\lambda$-states that would otherwise be highly unfavourable. This is
achievable by suitable choices of the pre-weighting factors $\eta_m$
featuring in equations~\ref{eq:Ze}--\ref{eq:ptilde}. Unfortunately
however, the choice of suitable $\eta_m$ cannot be determined directly
{\em a-priori}. For example, the weights that would yield a perfectly
flat sampled distribution (i.e. $p(\lambda_m)=$ constant), are given
by

\begin{equation}
\eta_m=\int_{\lambda_m}^1 \frac{d}{d\lambda}\ln Z(n,N,V,\lambda ) d\lambda
\label{eq:ideal}
\end{equation}
in which $Z(n,N,V,\lambda )$ is, of course, unknown. Nevertheless, it
turns out that it is possible to implement a straightforward iterative
scheme for obtaining suitable $\eta_m$.

\subsection{Determining the pre-weighting factors}

The goal is to determine values for the pre-weighting factors $\eta_m$
that allow the system to visit each $\lambda$-state with probabilities
of similar magnitude.  This may be accomplished by the following
procedure.

\begin{enumerate}

\item Choose a set of $M$ $\lambda$ values at which states are to be
placed. The exact value of $M$ is not too important at this stage,
although it should not be too small. A discussion of the optimal
choice for $M$ is given in section~\ref{sec:opt}. For most purposes,
however, between $5$ and $15$ states will suffice. For reasons that
will also be described later, these states should be distributed such that
their density increases with increasing $\lambda$.

\item Starting with all pre-weighting factors $\eta_m=0$, conduct a
relatively short simulation ($10^4$--$10^5$ Monte-Carlo sweeps), accumulating
the
histogram $p(\lambda_m)$. As a rule, the vast majority of counts will
be confined to the end state $m=M-1$. Provided, however, that the
adjacent state ($m=M-2$) is placed close to $\lambda=1.0$, the system will also
(albeit rarely) make some excursions to this state too.

\item Suppose now that counts were obtained for $p(\lambda_m)$ in
the $i$ states $M-i,M-i+1,..,M-1$. New estimates for $\eta_m$ are obtained
by linear extrapolation using the iterative formula:

\begin{eqnarray}
\eta_m^\prime =& \eta_m -\ln p(\lambda_m) & M-i \le m \le M-1\\ \nonumber
\eta_m^\prime =&
\eta_{M-i}+\frac{(\lambda_{M-i-1}-\lambda_{M-i})}{(\lambda_{M-i}-\lambda_{M-i+1})}
(\eta_{M-i}^\prime-\eta_{M-i+1}^\prime)  & 0 \le m < M-i
\addtocounter{equation}{-1}
\label{eq:guess}
\end{eqnarray}

\item Next perform another short simulation with the new pre-weighting
factors $\eta_m^\prime$.  In this second simulation, $p(\lambda_m)$
will receive counts in at least $i+1$ $\lambda$-states.

\end{enumerate}

Stages $3$ and $4$ are then simply repeated until weights are obtained
that allow the simulation to explore the entire $\lambda$ domain. This
iteration procedure can of course be easily automated if desired.
Once suitable weight factors are obtained, a longer simulation can be
performed to obtain the desired statistical accuracy for
$p(\lambda_m)$. The chemical potential is then obtained directly from
equation~\ref{eq:ptilrat}.

In some circumstances shortcuts to obtaining suitable weight factors
can also be exploited. For example, it is advantageous to carry out
the above procedure for a small system.  This entails little
computational effort, and assuming that the finite-size dependence of
\me\ is not too great, the weights acquired should serve for long runs
on larger systems.  Additionally, if it is desired to obtain \me\ for
a variety of chain lengths at a single density, weights determined for
a rather short chain can be extrapolated to longer chain lengths by
exploiting the approximate linear dependence of \me\ (and $\eta_m$) on
the chain length \cite{SZLEIFER}. Refinements to the extrapolation
procedure outlined above can also be envisaged. For instance, a
polynomial fit to all the weights already determined, might extend the
range of the extrapolation procedure, allowing two or more weights to
be determined per iterative cycle instead of just one.  In practice,
however, (and as is demonstrated in the following section), the above
procedure yields the weight factors surprisingly rapidly.

\section{Application to the Bond Fluctuation Model}

In order to test our method for measuring \me , we have implemented
an expanded ensemble simulation using the bond fluctuation model. This
is a lattice model for polymers in which polymer chains are
represented as $N$ monomer units, each occupying a unit-cell of a
simple cubic lattice. Successive monomers are connected by bonds, the
bond vector of which is allowed to assume one of $108$ distinct
values. The number of allowed bonds is dictated by prescribed
restrictions on the bond lengths and the necessity to prevent chains
from crossing one another. Aside from these restrictions however, no
explicit intra-molecular potential is incorporated. Excluded volume
effects are catered for by requiring mutual self avoidance of
monomers: no lattice site can be occupied simultaneously by two
monomers (unless, as is possible in the present case, one happens to
be a ghost monomer). For simulations in the thermal regime, an
attractive inter-molecular potential can be applied as required.
Polymer moves are facilitated either by a local monomer displacement
algorithm (as was the case in the present work), or alternatively by
reptation type moves.  Further details concerning the model can be
found in references \cite{BONDFL1,BONDFL2}.

Simulations were performed using a metropolis algorithm for both the
local monomer displacement algorithm and the $\lambda$-transitions
\cite{COMMENT2}. In the course of the simulations, a number of
different chain lengths and densities were studied in both the
thermal and athermal regimes.  We begin, however, by describing how the
weight factors $\eta_m$ were obtained for an athermal system of
polymers of length $N=40$ monomers in a volume $V=40^3$ at volume
fraction $\rho=0.4$.

A total of eight $\lambda$-states were employed, these being placed at
the $\lambda$ values shown in table~\ref{tab:etas}. It is not possible
to place a $\lambda$-state exactly at $\lambda=0$ since this causes a
computational error when the transition probabilities
(equation~\ref{eq:pacc}) are calculated. Nevertheless, the value of
$\lambda_0$ can be made extremely small, resulting in negligible
error.  For reasons made clear in subsection~\ref{sec:opt}, the
distribution of $\lambda$-states was chosen such that their density
increased with increasing $\lambda$ . The ratio of the number of displacement
moves per monomer to the number of attempted $\lambda$-transitions
was set at $4:1$ respectively.

The evolution of the iterative procedure by which the weights were
obtained, is also shown in table~\ref{tab:etas}.  Each run constitutes
one iterative cycle, in which a further weight factor was
determined. It is seen that for each run the extrapolated guess
(equation~\ref{eq:guess}) systematically overestimates the $\eta_m$
value to be determined, resulting in proportionally more weight for
the corresponding entry in $p(\lambda_m)$ on the ensuing run. However,
this overestimate is subsequently corrected on the following cycle,
when the weight factor is reduced with respect to the others. The
final weights are presented in the last column of
table~\ref{tab:etas}. Despite the need for several cycles to determine
the weight factors, the whole procedure was surprisingly rapid,
consuming only 30 minutes CPU time on an IBM RS/6000 workstation.

Having obtained suitable weight factors, a longer run of $2\times10^6$
Monte Carlo sweeps (MCS) was performed.  Figure~\ref{fig:plam40}(a)
shows the pre-weighted distribution $p(\lambda_m)$ obtained from this run,
while figure~\ref{fig:plam40}(b) shows the form of the reweighted
distribution $\tilde p(\lambda_m)\equiv\exp(-\eta_m)p(\lambda_m)$. The
associated chemical potential is $\me=61.4(1)$, corresponding to a
random insertion probability of $10^{-26}$.

With regard to figure~\ref{fig:plam40}(a), it should be pointed out
that the apparent noise in $p(\lambda_m)$ is illusory. The variations
in $p(\lambda_m)$ represent not statistical uncertainty (which in this
instance is much smaller than the variation), but rather deviations of
the weights from their `ideal' values as prescribed by
equation~\ref{eq:ideal}. Provided though that these deviations are not
so great as to hinder the random walk of the system between the
$\lambda$-states, there will be practically {\em no influence} of the
weight factors themselves on the statistical accuracy of \me , since
their effect is simply removed when the reweighted distribution
$\tilde p(\lambda_m)$ is formed.

To demonstrate that the method is effective for even longer chain
lengths and higher densities, we have studied an athermal system of
polymers of chain length $N=80$ at a volume fraction $\rho=0.5$,
contained in a volume $L=40^3$. The weight factors were determined
exactly as before, although as discussed in subsection~\ref{sec:opt},
more $\lambda$-states ($14$ in total) were required to prevent the
acceptance rate falling too low. The ratio of displacement moves per
monomer to attempted $\lambda$-transitions was set at $10:1$
respectively. Having determined suitable weight factors, a production
run of $4\times 10^6$ MCS was carried out, resulting in the
distributions shown in figure~\ref{fig:plam80}. The corresponding
estimate for the chemical potential is $\mu_{ex}=175.9(1)$, in good
agreement with that obtained previously under the same conditions
using thermodynamic integration \cite{MUELLER}.

Turning now to the thermal regime, we have employed a short-range
square well potential to study a system of volume $L=40^3$ and chain
length $N=40$ at a volume fraction $\rho=0.2$ and temperature
$T=2.0$. The procedure for determining the $\eta_m$ is similar to
before, although a complication arises in the fact that $\tilde p(\lambda_m)$
is no longer monotonic. The forms of $p(\lambda_m)$ and $\tilde
p(\lambda_m)$ are shown in figure~\ref{fig:plam_t}. The latter
distribution displays a broad minimum, manifesting the competition
between the positive athermal contribution to the chemical potential
(which dominates as $\lambda\rightarrow 1$) and the negative thermal
contribution (which dominates for small $\lambda$).  To obtain the
weight factors in this case it was necessary to start the
extrapolation procedure at both ends of the $\lambda$ range,
(i.e. $m=0$ and $m=M-1$) and work in towards the position of the
minimum.  The final complete set of weights was then obtained by
requiring continuity at the $\lambda$ value where the two sets
coincide (in effect by adding a constant to one set). The chemical
potential estimated from a run of $2\times10^6$ MCS is
$\mu_{ex}=-0.01(7)$. This very small value of \me\ indicates the close
proximity of the temperature to the $\Theta$-point of the model.

\subsection{Optimisation of the method}
\label{sec:opt}

In situations where very high statistical accuracy is required, long
runs can be necessary and a little effort to optimise the method may
pay dividends. Two interconnected factors influence the statistical
accuracy that can be realised for a given expenditure of computational
effort. Firstly there is the choice of the number of $\lambda$-states
$M$, and secondly there is the acceptance rate for
$\lambda$-transitions. These factors jointly control the magnitude of
the correlation time for the $\lambda$-transitions. To the extent that
$p(\lambda_m)$ can be regarded as flat, the system will execute a
one-dimensional random walk between the $\lambda$-states and the
correlation time will be given by:

\begin{equation}
\addtocounter{equation}{1}
\tau(M)\simeq \frac{M^2}{P_a(M)}
\label{eq:tau}
\end{equation}
where $\tau(M)$ is measured in units of transition attempts and
$P_a(M)$ is the acceptance rate for $\lambda$ transitions, which we
assume to be constant for each state. To appreciate how $P_a(M)$
depends on the number of $\lambda$-states $M$ (and their placing) it
is instructive to consider the transition probabilities for an
athermal system:

\begin{equation}
P_a(n\rightarrow m)=min(1,\exp [ N_o(\ln\lambda_m - \ln \lambda_n )+
\eta_m-\eta_n])
\label{eq:pacca}
\end{equation}
In addition to ensuring that $p(\lambda_m)$ is approximately flat, the
r\^{o}le of the weights in this equation is to partially compensate
for positive changes in the exponent associated with transitions to
smaller $\lambda$ values. The extent to which they are successful in
this regard (and hence the size of the acceptance rate) depends on the
degree of overlap between the distributions of $N_o(\lambda_m)$ and
$N_o(\lambda_n)$. In order to ensure a sufficiently high acceptance
rate, sufficient $\lambda$-states must be employed to guarantee an
appreciable overlap between the distributions. Moreover, to maintain
an approximately constant acceptance rate for all states it is
necessary to increase their density at higher $\lambda$ values. This
is illustrated in figure~\ref{fig:nover} where the distribution
function $N_o(\lambda)$ is plotted for a variety of fixed $\lambda$
values. The distributions shown have approximately equal overlap
although the $\lambda$-states are placed closer together for larger $\lambda$.

The above considerations illustrate the compromise to be struck in
minimising the correlation time: if $M$ is too large, $\tau$ will be
large and accordingly it will be difficult to accumulate uncorrelated
statistics. Similarly, if $M$ is too small, a low acceptance rate will
ensue and consequently $\tau$ will again be large. However, it should
be stressed that the correlation time is not the sole factor governing
the optimum choice of $M$. For a given length of run, the statistical
quality of $p(\lambda_m)$ may be improved by choosing a smaller value
of $M$, even at the expense of a larger correlation time. If the
sampling frequency is chosen commensurate with the correlation time
(equation~\ref{eq:tau}), a fraction $2/M$ of the total number of
samples will fall into the two extrema states $p(\lambda_0)$ and
$p(\lambda_{M-1})$ used to calculate \me . Now, since the statistical
error on \me\ is dependent on the number of entries in these end
states, it follows that the optimal choice of $M$ is given, not by the
minimum in $\tau(M)$, but by the minimum in $M^3/P_a(M)$. In
figure~\ref{fig:pacc} we plot this function for a system of athermal
polymers with $N=40$ and $\rho=0.2$. A minimum obtains for $M=6$,
corresponding to an acceptance rate $P_a\approx 20\%$. Thus it would
appear that in the present method, the optimal acceptance rate can be
considerably less than the value of $50\%$ or so used for normal
particle displacements.


Finally we address briefly the question of the optimal choice of the
ratio of local displacement moves to $\lambda$-transitions. Clearly
the issue is again one of efficiency. If too few local moves are
attempted between transition attempts, then the number of overlaps
$N_o$ will systematically deviate from the canonical average
$<N(\lambda)>$ and the acceptance rate may suffer. On the other hand,
if the ratio of local moves to transition attempts is too high, a
longer run will be necessary to obtain good statistics in
$p(\lambda_m)$.  For the present model, we find that following a
$\lambda$ transition, between $4$ $(N=40)$ and $10$ $(N=80)$ local
moves attempts per monomer are sufficient to decorrelate the number of
overlaps. We note also in this context, that relatively few local
moves are required to decorrelate the spatial position of the ghost
polymer.  This is because the ghost polymer frequently samples
$\lambda$-states in which it completely decouples from the normal
polymers, thereby allowing it to diffuse around the system with
relative ease.  This latter feature of the method is in favourable
contrast to the thermodynamic integration method \cite{MUELLER}, in
which several hundred trial displacements are required for a ghost
polymer with fixed small $\lambda$ to diffuse away from its starting
point.

\subsection{Finite-size corrections to the chemical potential}

Measurements of the chemical potential in finite-size systems are
known to be relatively strongly effected by finite-size
effects. Indeed for a system of $n$ particles, the leading corrections
vary like $n^{-1}$. Recently, Siepmann {\em et al} have derived an
expression for the leading finite-size dependence of \me\ in terms of
density derivatives of the pressure \cite{SIEPMANN}

\begin{equation}
\Delta \mu_{ex}( n ) =\frac{1}{2n} \left ( \frac{\partial P}{\partial
\rho }\right ) \left [ 1-k_bT\left (\frac{\partial \rho}{\partial
P}\right ) - \rho k_bT\frac{(\partial^2 P/\partial\rho^2)}{(\partial
P/\partial \rho )^2} \right ] + O(n^{-2})
\label{eq:muN}
\end{equation}
where $\rho=n/V$ is the polymer number density. Provided that the
equation of state is known independently, this equation
permits a calculation of the finite-size dependence of the chemical
potential, thereby facilitating an extrapolation to the thermodynamic
limit.

As a test of the precision and accuracy of our method, we have
attempted to measure the finite-size dependence of \me\ for our model
and compare it with equation~\ref{eq:muN}. To this end we have
simulated systems of athermal polymer chains of length $N=40$ at
volume fraction $\rho=0.2$ for system sizes $L=40, 47$ and $60$. In
accord with the findings of the previous subsection, 6
$\lambda$-states were utilised. Runs comprising $10^7$ MCS were
performed, and the $\lambda$ value was sampled every $180$ transition
attempts. For each system size studied, $16$ independent runs were
carried out in order to test the statistical independence of the data
and assign errors to the results.

The resulting estimates for the excess chemical potential are
$\mu_{ex}(40)=19.78(2), \mu_{ex}(47)=19.69(2), \mu_{ex}(60)=19.59(2)$,
where the value for $L=47$ has been corrected for the fact that
$\rho=0.2$ does not correspond to a whole number of
polymers. Figure~\ref{fig:fss} shows how these estimates compare with
equation~\ref{eq:muN}. The broken line represents the predicted
finite-size shift in \me\ obtained by feeding the equation of state
data of M\"{u}ller and Paul \cite{MUELLER} into
equation~\ref{eq:muN}. Assuming an infinite-volume chemical potential
of $\me (\infty)=19.50$, the measured finite-size shift in \me\ is
consistent to within error with the prediction of
equation~\ref{eq:muN}.

\section{Conclusions}

In summary, we have presented an efficient method for measuring the
chemical potential of polymeric systems that is effective at both
large chain lengths and high densities. Although some preliminary
effort is required to find suitable pre-weighting factors, this is not
excessively time consuming and is generously rewarded by the very high
accuracy that can subsequently be attained. In contrast to the Chain
Increment method \cite{KUMAR1}, the estimate of \me\ derives from a
single simulation and is in principle exact. The present method should
therefore provide a more accurate alternative in applications such as
determining the polymer-solvent coexistence curve of long chain
molecules (where configurational bias Gibbs ensemble simulations are
not feasible). Finally we remark that the present approach is not
limited to homopolymer chains, and should also facilitate measurements
of \me\ for copolymers as well as more complex polymer architectures
such as branched or ring structures.

\subsection*{Acknowledgements}

The authors have benefitted from helpful discussions with G.R. Smith,
K. Binder, P. Rehberg and W. Janke.  Partial support from the Deutsche
Forschungsgemeinschaft (DFG) under grant number BT314/3-2 is
gratefully acknowledged.

\newpage

\begin{figure}[h]
\vspace*{0.5 in}
\caption{(a) The pre-weighted distribution $p(\lambda_m)$ for a system of
athermal polymers of chain length $N=40$ and volume fraction $\rho=0.4$,
contained in a volume $V=40^3$. (b) The reweighted distribution
$\tilde p(\lambda_m)=\exp(-\eta_m)p(\lambda_m)$
expressed on a logarithmic scale.}
\label{fig:plam40}
\end{figure}

\begin{figure}[h]
\vspace*{0.5 in}
\caption{(a) The pre-weighted distribution $p(\lambda_m)$ for a system of
athermal polymers of chain length $N=80$ and volume fraction $\rho=0.5$,
contained in a volume $V=40^3$. (b) The reweighted distribution
$\tilde p(\lambda_m)=\exp(-\eta_m)p(\lambda_m)$ expressed on a logarithmic
scale.}
\label{fig:plam80}
\end{figure}

\begin{figure}[h]
\vspace*{0.5 in}
\caption{(a) The pre-weighted distribution $p(\lambda_m)$ for a system of
thermal polymers of chain length $N=40$, $\rho=0.2$ at temperature $T=2.0$,
contained in a volume $V=40^3$. (b)  The reweighted distribution
$\tilde p(\lambda_m)=\exp(-\eta_m)p(\lambda_m)$ expressed on a logarithmic
scale.}
\label{fig:plam_t}
\end{figure}

\begin{figure}[h]
\vspace*{0.5 in}
\caption{Distributions of the number of
overlaps between ghost monomers and normal monomers at various fixed
$\lambda$.}
\label{fig:nover}
\end{figure}

\begin{figure}[h]
\vspace*{0.5 in}
\caption{The measured form of the function $M^3/P_a(M)$}
\label{fig:pacc}
\end{figure}

\begin{figure}[h]
\vspace*{0.5 in}

\caption{The measured value of the chemical potential for a system of
athermal polymers of chain length $N=40$ at volume fraction $\rho=0.2$
for system sizes $L=40, 47, 60$. A constant corresponding to the
infinite volume estimate $\mu_{ex}=19.50$ has been subtracted from the
data. Also shown (broken curve) is the predicted finite-size
dependence of \protect\me\ following from
equation~\protect\ref{eq:muN} utilising predetermined equation of
state data \protect\cite{MUELLER}.}

\label{fig:fss}
\end{figure}

\begin{table}[h]

\begin{center}
\begin{tabular}{|l|c|c||c|c||c|c|} \hline
&\multicolumn{2}{|c||}{Run 1}&\multicolumn{2}{|c||}{Run 2}
&\multicolumn{2}{|c|}{Run 3}\\ \hline \hline
\multicolumn{1}{|c|}{$\lambda_m$} & $\eta_m$ & $p(\lambda_m$)& $\eta_m$
&  $p(\lambda_m)$& $\eta_m$& $p(\lambda_m)$  \\ \hline
$\lambda_0=0.00001$&0.00 &0.00  & 12.60  & 0.000 & 24.28 & 0.000\\
$\lambda_1=0.35$   &0.00 &0.00  & 12.60  & 0.000 & 24.28 & 0.000\\
$\lambda_2=0.55$   &0.00 &0.00  & 12.60  & 0.000 & 24.28 & 0.000 \\
$\lambda_3=0.7$    &0.00 &0.00  & 12.60  & 0.000 & 24.28 & 0.000 \\
$\lambda_4=0.8$    &0.00 &0.00  & 12.60  & 0.000 & 24.28 & 0.567 \\
$\lambda_5=0.9$   &0.00 &0.00  & 12.60  & 0.423 & 12.18 & 0.168 \\
$\lambda_6=0.97$   &0.00 &0.022 & 3.79   & 0.299 & 3.72  & 0.142 \\
$\lambda_7=1.0$    &0.00 &0.977 & 0.00   & 0.277 & 0.00  & 0.113 \\
\hline
\end{tabular}

\vspace*{1cm}

\begin{tabular}{|l|c|c||c|c||c|c||c|} \hline
&\multicolumn{2}{|c||}{Run 4}&\multicolumn{2}{|c||}{Run 5}
&\multicolumn{2}{|c||}{Run 6}&\\ \hline \hline
\multicolumn{1}{|c|}{$\lambda_m$} & $\eta_m$ & $p(\lambda_m$)& $\eta_m$
&  $p(\lambda_m)$& $\eta_m$ & $p(\lambda_m)$&$\eta_m$\\ \hline
$\lambda_0=0.00001$& 33.63 & 0.00  & 47.13 & 0.00  & 57.78 & 0.009& 61.02\\
$\lambda_1=0.35$   & 33.63 & 0.00  & 47.13 & 0.00  & 57.78 & 0.673& 53.67\\
$\lambda_2=0.55$   & 33.63 & 0.00  & 47.13 & 0.656 & 45.03 & 0.060& 43.90\\
$\lambda_3=0.7$    & 33.63 & 0.423 & 32.77 & 0.052 & 33.21 & 0.063& 32.50\\
$\lambda_4=0.8$    & 22.75 & 0.113 & 23.21 & 0.064 & 23.44 & 0.042& 23.14\\
$\lambda_5=0.9$   & 11.86 & 0.133 & 12.16 & 0.066 & 12.37 & 0.036& 12.21\\
$\lambda_6=0.97$   & 3.57  & 0.153 & 3.73  & 0.079 & 3.75  & 0.034& 3.64\\
$\lambda_7=1.0$    & 0.00  & 0.177 & 0.00  & 0.082 & 0.00  & 0.031& 0.00\\
\hline
\end{tabular}
\end{center}

\caption {The weight determination procedure described in the text for
N=40, $\rho=0.2$. All runs are 50000 MCS.}

\label{tab:etas}
\end{table}


\begin{thebibliography}{20}
\bibitem{COMMENT2} Care must be exercised to ensure that detailed balance
is satisfied when the system is in state $\lambda_0$ or
$\lambda_{M-1}$, there being only one choice for the new trial
$\lambda$-state.
\end{thebibliography}
\end{document}